\documentstyle[preprint,aps]{revtex}
\begin{document}
\draft
\preprint{\hbox to \hsize{\hfil\vtop{\hbox{IASSNS-HEP-96/75}}}}
\title{Response to the Comment by G. Emch on Projective Group\\
Representations in Quaternionic Hilbert Space\\
}
\author{ Stephen L. Adler\\
}

\address{
Institute for Advanced Study\\
Princeton, NJ 08540\\ 
\phantom{$\sum$}
{\tt Running head:~~Quarternionic projective group
representations}\\}

\maketitle

\begin{abstract}
We discuss the differing definitions of complex and quaternionic 
projective group  
representations employed by us and by Emch.  The definition of Emch 
(termed here a strong projective representation) is 
too restrictive to accommodate quaternionic Hilbert space embeddings 
of complex projective representations. Our definition (termed here a weak 
projective representation) encompasses 
such embeddings, and leads to a detailed theory 
of quaternionic, as well as complex, projective group representations.  
\end{abstract}

\section*{\bf I.~~Preliminaries not involving group structure\\}

Before turning to a discussion of what is an appropriate definition of a
quaternionic projective group representation, we first address several 
issues that do not involve the notion of a {\it group} of symmetries.  
We follow throughout the Dirac notation used in our recent book [1], 
in which linear operators in Hilbert space act on ket states from the 
left and on bra states from the right, as in ${\cal O} |f \rangle$
and $\langle f|\cal O$, while quaternionic scalars in Hilbert 
space act on ket states from the right and on bra states from the left, as in 
$|f \rangle \omega$ and $\omega \langle f|$.

We begin by recalling the statement (see Sec. 2.3 of Ref.~[1]) 
of the quaternionic extension of                                   
Wigner's theorem, which gives the Hilbert space representation of an 
individual 
symmetry in quantum mechanics.  Physical states in quaternionic quantum  
mechanics are in one-to-one correspondence with unit rays of the form
$|{\bf f} \rangle=\{ |f\rangle \omega \}$, with $|f\rangle$ a unit normalized 
Hilbert space vector and $\omega$ a quaternionic phase of unit magnitude.  A
symmetry operation $\cal S$ is a mapping of the unit rays $|{\bf f}\rangle$ 
onto images $|{\bf f}^{\prime} \rangle$, which preserves all transition 
probabilities,
\begin{eqnarray}
{\cal S} |{\bf f}\rangle &=& |{\bf f}^{\prime} \rangle \nonumber\\
|\langle {\bf f}^{\prime}|{\bf g}^{\prime} \rangle|&=&
|\langle {\bf f} | {\bf g} \rangle |. 
\label{one}
\end{eqnarray}
Wigner's theorem, as extended to quaternionic Hilbert space, asserts that 
by an appropriate $\cal S$-dependent choice of ray representatives for the 
states, the mapping $\cal S$ can always be represented 
(in Hilbert spaces of dimension greater 
than 2) by a unitary transformation $U_{\cal S}$ on the state vectors, 
so that 
\begin{equation}
|f^{\prime}\rangle = U_{\cal S}|f \rangle~. 
\label{two}
\end{equation}
Conversely, any unitary transformation of the form of Eq.~(2) clearly implies 
the preservation of transition probabilities, as in Eq.~(1).  When only one  
symmetry transformation is involved, the issue of projective 
representations does not enter, since Wigner's theorem asserts that this 
transformation can be given a unitary representation on appropriate 
ray representative states in Hilbert space.  The issue of projective 
representations arises only when we are dealing with two (or more) symmetry 
transformations, in which case the ray representative choices which reduce 
the first symmetry transformation 
to unitary form may not be compatible with the ray representative 
choices which reduces a second symmetry transformation to unitary form.  
Thus we disagree with Emch's statement, in the semifinal paragraph of his  
Comment, that Wigner's theorem (which he notes is a form of the first 
fundamental theorem of projective geometry) may be dependent on the 
definition adopted for quaternionic projective group representations.  

In the first section of his Comment, Emch proves a Proposition stating that  
if an operator $\cal O$ commutes with all of the projectors 
$|f\rangle \langle f| $ of a quaternionic Hilbert space of dimension 2 
or greater, then $\cal O$ must be   
a real multiple of the unit operator $1$ in Hilbert space.  When $\cal O$  
is further restricted to be a unitary operator (as obtained from a symmetry 
transformation via the Wigner theorem), the real multiple is further 
restricted to be $\pm 1$.  Since we will refer to this result in the next 
section, let us give an alternative proof, based on the 
spectral representation of a general unitary operator $U$ in quaternionic 
Hilbert space, 
\begin{equation}
U=\sum_{\ell}|u_{\ell}\rangle e^{i \theta_{\ell}} \langle u_{\ell}|~,
~~0 \le \theta_{\ell} \le \pi~,
\label{three} 
\end{equation}
in which the sum over $\ell$ spans a complete set of orthonormal 
eigenstates of $U$.
Let us focus on a two state subspace spanned by $|u_1\rangle$ and 
$|u_2\rangle$, and construct the projector $P=|\Phi\rangle \langle\Phi|$,  
with 
\begin{eqnarray}
|\Phi\rangle&=&|u_1\rangle + |u_2\rangle \omega~,\nonumber\\
\overline{\omega}&=&-\omega~,~~\omega=\omega_{\alpha} +j \omega_{\beta}~,~~
\omega_{\alpha}\omega_{\beta} \ne 0 ~,
\label{four}
\end{eqnarray}
where $\omega_{\alpha,\beta}$ are symplectic components lying in the complex 
subalgebra of the quaternions spanned by $1$ and $i$.  Then the
projector $P$ is given by 
$$
P=|u_1\rangle \langle u_1|+|u_2 \rangle \langle u_2|
+|u_2 \rangle \omega \langle u_1|-|u_1 \rangle \omega \langle
u_2|~,\eqno(5a)
$$
and the part of $U$ lying in the $|u_{1,2}\rangle$ subspace is 
$$
U_{1,2}=|u_1\rangle e^{i \theta_1} \langle u_1| 
+|u_2\rangle e^{i \theta_2} \langle u_2| ~.\eqno(5b)
$$
The commutator of $U$ and $P$ is then given by 
\setcounter{equation}{5}
\begin{equation}
[U,P]=[U_{1,2},P]=
|u_2\rangle (e^{i \theta_2} \omega -\omega e^{i \theta_1})\langle u_1|  
-|u_1\rangle (e^{i \theta_1} \omega -\omega e^{i \theta_2})\langle u_2|~,  
\label{six}
\end{equation}
which vanishes only if $e^{i \theta_1}=e^{i \theta_2}$ (from equating to 
zero the coefficient of $\omega_{\alpha}$) and $e^{i \theta_1} = 
e^{-i \theta_2}$ (from equating to zero the coefficient of $\omega_{\beta}$).
Since $0 \le \theta_{1,2} \le \pi$, this requires either $\theta_1=
\theta_2=0$ or $\theta_1=\theta_2=\pi$.  Repeating the argument for each 
dimension 2 subspace in turn, we learn that $U=\pm 1$.  Note that in a 
complex Hilbert space, the analogous argument shows only that 
$e^{i \theta_1}=e^{i \theta_2}$, from which we conclude 
(again by repeating the 
argument for each dimension 2 subspace in turn) that $U=e^{i\theta}$, 
which commutes with all projectors because any complex number is 
a $c$-number in complex Hilbert space.

Clearly, the argument just given involves only elementary properties of 
the projectors in Hilbert space, and makes no reference to the notion of 
a group of symmetries.  The same is true of the proposition given in Sec. I 
of Emch's Comment.  Since Schur's Lemma ordinarily describes the restrictions 
on an operator that commutes with the representation matrices of an 
irreducible group representation, 
and since the projectors in Hilbert space do not form a group (they 
are not invertible and the product of two different projectors is not 
a projector), 
it is a misnomer to describe Emch's 
Proposition, or the corollary given here, as a ``quaternionic Schur's 
lemma''. 
In addition to disagreeing with Emch's terminology, we also 
disagree with his statement, in the second paragraph of Sec. III of 
his Comment,  
that the analysis leading to his Proposition is 
dependent on the definition adopted for quaternionic projective group 
representations; in fact, the notion of a group of symmetries does not enter 
into either his analysis, or the corollary for unitary matrices proved here.  

\section*{II.~~How should one define quaternionic projective
group representations?}

Let us now address the central question of how one should generalize to 
quaternionic Hilbert space the notion of a projective group representation. 
We begin by reviewing how projective group representations arise in complex  
Hilbert space.  Let $\cal G$ be a symmetry group composed of abstract 
elements $a$ with group multiplication $ab$.  By Wigner's theorem, each 
group element is represented, after an $a$-dependent choice of ray 
representatives, by a unitary operator $U_a$ acting on the states of 
Hilbert space.  In the simplest case, in which the $U_a$ are said to 
form a vector representation, the $U$'s obey a multiplication law 
isomorphic to that of the corresponding abstract group elements, 
\begin{equation}
U_aU_b=U_{ab}~.
\label{seven}
\end{equation}
However, when the complex rephasings of the states used in Wigner's 
theorem are taken into account, there exists the more general possibility  
that for any state $|f\rangle$, the states $U_aU_b|f\rangle $ and 
$U_{ab}|f \rangle$ are not equal, but rather differ from one another by a 
change of ray representative, i.e., 
\begin{equation}
U_aU_b|f\rangle = U_{ab}|f \rangle e^{i \phi(a,b;f)}  ~.
\label{eight}
\end{equation}
Corresponding to Eq.~(8), there are two possible definitions of a projective 
representation in complex Hilbert space:\hfill\break

{\it Definition (1)}  In a {\it weak} projective representation, the 
multiplication 
law of the $U$'s obeys Eq.~(8) on one complete set of states $\{ |f\rangle\}$.
This suffices, by superposition, to determine the multiplication law of the 
$U$'s on all states. \hfill \break

{\it Definiton (2)} In a {\it strong} projective representation, the  
multiplication law of the $U$'s obeys Eq.~(8) on all states in Hilbert 
space. In this case, we can easily prove that the phases $\phi(a,b;f)$ are 
independent of the state label $f$.  To see this, let us define 
$V_{ab}=U_{ab}^{-1}U_aU_b$; then Eq.~(8) implies that 
\begin{equation}
V_{ab}|f\rangle=|f \rangle e^{i \phi(a,b;f)}~, 
\label{nine}
\end{equation}
which immediately implies that $V_{ab}$ commutes with the projector 
$|f\rangle \langle f|$, for all states $|f\rangle$ in Hilbert space.  
But invoking the complex Hilbert space specialization of the result of 
the preceding section, we learn that $V_{ab}$ must be a $c$-number, 
$V_{ab}=e^{i\phi(a,b)}$.  This is the customary definition of a projective 
representation in complex Hilbert space, and is well known to have nontrivial 
realizations. 

Let us now turn to the question of how to define projective representations 
in quaternionic Hilbert space.  Emch choses as his generalization the 
strong definition given above, which by the reasoning following Eq.(9), 
and the quaternionic result of Sec. 1, implies that $V_{ab}=(-1)^{n_{a,b}}$, 
with $n_{a,b}$ an integer that can depend in general on $a$ and $b$.
In other words, {\it the only strong quaternionic projective representations 
are real projective representations}.  

The problem with adopting the strong definition, however, is that it excludes 
from consideration as a quaternionic projective representation the 
embedding into quaternionic Hilbert space of a nontrivial complex projective 
representation realized on a complex Hilbert space.  Thus, potentially 
interesting structure is lost.  To avoid this problem, Ref. [1] adopts as 
the quaternionic generalization of the notion of a projective representation 
the weak definition given above, which in quaternionic Hilbert space 
states that
\begin{equation}
U_aU_b|f\rangle=U_{ab}|f\rangle \omega_{a,b}~,~~|\omega_{a,b}|=1~ 
\label{ten}
\end{equation}
for one particular complete set of states $\{|f\rangle \}$.  As discussed 
in Ref.~1, Eq.~(10) can also be rewritten in the operator form 
$$
U_aU_b=U_{ab}\Omega(a,b) ~,
\eqno(11a)
$$
with 
$$
\Omega(a,b)=\sum_f |f\rangle \omega(a,b;f)\langle f|~.
\eqno(11b)
$$
Since the operator $\Omega$ depends on the particular complete set 
of states on which the projective phases are given, a more complete notation 
(not employed in Ref.~1) would in fact be $\Omega(a,b;\{ |f\rangle\})$.  
Using the result of an analysis [2] of the associativity condition for 
weak quaternionic projective representations, Tao and Millard [3] have 
recently given a beautiful complete structural classification theorem for 
weak quaternionic projective representations.  The complex  
specialization of their Corollary 2, incidentally, 
states that in a complex Hilbert space, 
the weak definition of a projective representation implies the strong one.

Can the weak definition of a quaternionic projective representation 
be weakened even further, 
by using a {\it different} complete set of states $\{ |f\rangle\}$ to 
specify the projective phases for each pair of group elements $a$ and $b$ [4]?
In this case, the operator $\Omega$ takes the form $\Omega(a,b; \{ |f\rangle \}
_{a,b})$.  However, since any unitary operator is diagonalizable on 
some complete set of states, this further weakening allows an
arbitrary specification 
of $\Omega$ for each $a,b$, and any relationship of the unitary 
representation to the underlying group structure is lost.

\section*{III.~~ Discussion}
We conclude that the difference between our analysis and that of Emch 
is traceable to what I have here termed the difference between a {\it strong} 
and a {\it weak} definition of projective representation.  The strong 
definition 
is the customary one in complex Hilbert space, but it 
excludes potentially interesting structure when applied to quaternionic 
Hilbert space.  Since the weak definition leads to a detailed theory 
[1, 2, 3] of projective group representations in quaternionic Hilbert 
space, and since it implies [3] the strong definition in complex Hilbert 
space, the weak definition is in fact the more appropriate one in both 
complex and quaternionic Hilbert spaces.

\acknowledgments
This work was supported in part by the Department of Energy under
Grant \#DE--FG02--90ER40542. I wish to thank A.C. Millard and T. Tao for 
informative conversations, and to acknowledge the hospitality of the 
Aspen Center for Physics, where this work was done.

\end{document}